\documentclass[preprint,12pt]{elsarticle}

\usepackage{amssymb}
\usepackage{xcolor}
\usepackage{float}
\usepackage{graphics}

\journal{Physica B: Condensed Matter}

\begin{document}

\begin{frontmatter}

\title{Ground-state properties of the one-dimensional Hubbard model with pairing potential}

\author[lpt,hongik]{Myung-Hoon Chung}
\ead{mhchung@hongik.ac.kr}

\author[lpensl]{Edmond Orignac}
\ead{Edmond.Orignac@ens-lyon.fr}

\author[lpt]{Didier Poilblanc}
\ead{Didier.Poilblanc@irsamc.ups-tlse.fr}

\author[lpt]{Sylvain Capponi\corref{cor}}
\ead{capponi@irsamc.ups-tlse.fr}

\address[lpt]{Laboratoire de Physique Th\'{e}orique, IRSAMC, Universit\'{e} de Toulouse, CNRS, UPS, France}
\address[hongik]{College of Science and Technology, Hongik University, Sejong 339-701, Korea}
\address[lpensl]{Univ Lyon, Ens de Lyon, Univ Claude Bernard, CNRS, Laboratoire de Physique, F-69342 Lyon, France}
\cortext[cor]{Corresponding author}

\date{\today}

\begin{abstract}
We consider a modification of the one-dimensional Hubbard model by
including an external pairing potential. Guided by analytic bosonization results, we quantitatively determine the
grand-canonical zero-temperature phase diagram using both finite
and infinite density matrix renormalization group algorithm based
on the formalism of matrix product states and  matrix product
operator, respectively. By computing various local quantities as
well as the half-system entanglement, we are able to distinguish
between Mott, metallic and superconducting phases. We point out
the compressible nature of the Mott phase and the fully gapped
nature of the many-body spectrum of the superconducting phase, in
the presence of explicit U(1)-charge symmetry breaking.
\end{abstract}

\begin{keyword}

density matrix renormalization group \sep Hubbard model \sep
entanglement entropy \sep bosonization

\PACS 71.27.+a \sep 02.70.-c \sep 03.67.-a

\end{keyword}

\end{frontmatter}

\section{Introduction}

Long-range quantum correlations often fully
characterize the nature of a quantum phase in many-particle systems.
An abrupt change of correlations typically occurs at
quantum phase transitions \cite{Sachdev}. As a quantitative measure
of quantum correlations, the entanglement entropy plays a central role, potentially able to
signal quantum phase transitions or to characterize critical gapless phases \cite{Cha}.

To define the entanglement entropy \cite{Eisert}, one bipartites the quantum
system into two parts $A$ and $B$. One then introduces a density matrix
$\rho = |\Psi \rangle \langle \Psi |$ of a pure quantum state
$|\Psi \rangle$, and obtains the reduced density matrix
$\rho_{A}=\mbox{Tr}_{B}\rho$ by tracing out the subsystem $B$. The
entanglement entropy is the von Neumann entropy, which is given by
$S_{A} = -\mbox{Tr}(\rho_{A}\ln \rho_{A})$. Since we will consider zero-temperature properties in the following, we will compute entanglement properties using the ground-state as wavefunction $|\Psi \rangle$.

In order to find the ground state in quantum many-particle
systems, the density matrix renormalization group~\cite{White} (DMRG) method is
suitable, especially in one dimension. The connection between DMRG and tensor networks was
first recognized by the quantum information community
\cite{Ostlund}. A detailed reformulation of DMRG in terms of
matrix product states (MPS) was reviewed by Schollw\"{o}ck
\cite{Schollwoeck}. The generalization of MPS to handle
two-dimensional systems was carried out, and the projected
entangled pair state (PEPS) was introduced \cite{Verstraete}. For
critical systems, the multi-scale entanglement renormalization
ansatz (MERA) is useful \cite{Vidal1}. When a Hamiltonian has
translational invariance, we can use the so-called infinite DMRG
(iDMRG) \cite{McCulloch}, in which we assume that the matrices in
the MPS are identical. Without entering into too much details, let us simply mention that we update
a few matrices and environments to converge to the ground-state $|\Psi_{0} \rangle$ with the iDMRG method.

One motivation of introducing a pairing term is to model condensed matter quasi-1D electronic quantum wires weakly coupled to a thin (3D) supercconductor sheet. Such a set-up has been proposed by Kitaev~\cite{Kitaev2001} to realize boundary Majorana fermions by proximity effect with a p-wave superconductor. Our proposal involves a singlet (s-wave or d-wave) superconductor that could be realized using e.g. a (quasi-2D) high-Tc superconductor.

Also, since atoms with an odd number of neutrons \cite{Parsons} are
fermions and thus obey the same statistical rules as electrons,
some cold atoms \cite{Cheuk} in an optical lattice can mimic the
behaviors of electrons in a real solid material. In cold atom
experiments, quantum gas microscope \cite{Bakr} was used to create
and image an antiferromagnet, a phase in which one atom occupies
each lattice site, and the spins of neighboring atoms point in
opposite directions \cite{Mazurenko}. This  provides us with a motivation to consider the
singlet state of neighboring atoms with the pairing interaction.

Since experimentalists can design ultracold many-fermion systems loaded on
quasi one-dimensional optical lattices \cite{Schreiber}, the
one-dimensional fermion Hubbard model has become a
physical reality. Quite interestingly, the attractive Hubbard model which is the simplest model to describe
pairing and superconductivity in a fermionic system can also be realized~\cite{Mitra2018}. In our case,
the {\it repulsive} Hubbard model with an additional pairing potential, providing a tendency to form nearest-neighbor singlet pairs,
could also be realized in fermionic systems, for instance by proximity effect with a singlet superconductor.

The purpose of this paper is to compute the ground-state phase diagram
of the one-dimensional Hubbard model with a (singlet) pairing
potential. To do so, guided by analytic bosonization results, we apply two standard numerical algorithms (finite and infinite DMRG)
separately to optimize the matrices in the MPS. The results obtained by both methods are consistent with
each other. By changing the chemical potential, a quantum phase transition occurs between gapless and gapped phases, as can be measured from the scaling of the entanglement
entropy.

\section{Model and analytical treatments}\label{sec:model}

\subsection{Model and some exact transformations}

We consider a simple generalization of the Hubbard model on a
one-dimensional (1D) lattice~:
\begin{eqnarray}
H &=& -t\sum_{\langle i j
\rangle}(c^{\dagger}_{i\uparrow}c_{j\uparrow} +
c^{\dagger}_{j\uparrow}c_{i\uparrow} +
c^{\dagger}_{i\downarrow}c_{j\downarrow} +
c^{\dagger}_{j\downarrow}c_{i\downarrow}) \nonumber \\
& & +U\sum_{i}(n_{i\uparrow}-\frac{1}{2})(n_{i\downarrow}-\frac{1}{2})
-\mu \sum_{i}(n_{i\uparrow}+n_{i\downarrow})  \nonumber \\
& & -\Delta\sum_{\langle i j
\rangle}(c^{\dagger}_{i\uparrow}c^{\dagger}_{j\downarrow} +
c^{\dagger}_{j\uparrow}c^{\dagger}_{i\downarrow} +
c_{i\downarrow}c_{j\uparrow} +
c_{j\downarrow}c_{i\uparrow}),
\label{eq:H}
\end{eqnarray}
where $c$ and $c^\dagger$ are the usual spin-1/2 fermion annihilation and creation operators, $n_{i\sigma}=c^\dagger_{i\sigma}c_{i\sigma}$ is the local spin-resolved density and
$\langle i j \rangle$ stands for nearest neighboring sites on the 1D chain.
We fix the hopping
strength $t=1$ (as unit of energy) and vary the other three parameters:
the on-site Coulomb repulsion $U$, the chemical potential $\mu$,
and the pairing strength $\Delta$. The role of the chemical
potential is to control the average number of fermions in the system. Note that the (bond) singlet pairing potential does not conserve the particle number so that the model only has SU(2) spin symmetry.
Physically, such a potential may account for the proximity effect of a nearby singlet superconductor.
Without pairing potential, i.e. for $\Delta=0$, we recover the standard
one-dimensional (repulsive) Hubbard model which will be used for benchmark calculations as it is exactly solvable
\cite{Yang,hubbard_book}.

Let us make some additional remarks about the symmetries of this model. Half-filling will correspond to $\mu=0$ obviously and the phase diagram will be symmetric under $\mu \leftrightarrow -\mu$. Moreover, when $\mu=0$, applying a particle-hole symmetry only on odd sites ($d_{2i,\sigma}=c_{2i,\sigma}$ and $d_{2i+1,\sigma}=c^\dagger_{2i+1,-\sigma}$) amounts to exchanging the hamiltonian parameters as $(t,U,\mu=0,\Delta) \leftrightarrow (\Delta,U,\mu=0,t)$, {\it i.e.} exchanging the role of $t$ and $\Delta$.

In the non-interacting case ($U=0$), the model is quadratic so that it can be
diagonalized in Fourier space using a Bogoliubov transformation to get
\begin{equation}
  H_0 = \sum_k E_k (\alpha^\dagger_k \alpha_k + \beta^\dagger_k \beta_k)
\end{equation}
with a dispersion $E_k= \pm \sqrt{(\varepsilon_k-\mu)^2 + (\Delta \cos k)^2}$ where $\varepsilon_k = -2t \cos k$ is the tight-binding dispersion. In particular, for a generic filling ({\it i.e.} a generic $\mu$ value), we have a one-dimensional superconductor with a finite gap~\footnote{In the many-body spectrum, the ground-state is unique and there is a finite gap $2|\mu|$ for the first excitation.}. Indeed, there is no U(1) symmetry breaking in this model (since the particle number conservation is \emph{explicitly} broken) and, hence, no emergent zero-energy Goldstone modes. On general grounds, we expect that this superconducting phase will persists in some range of the phase diagram, even in the presence of a finite repulsive $U$. Note also that, in this gapped superconducting phase, the compressibility is finite though.

We now transform the Hamiltonian in a convenient form. Using the Bogoliubov rotation
\begin{eqnarray}
  \label{eq:bogoliubov}
  c_{j\uparrow} = \cos \frac \theta 2 f_{j\uparrow} - \sin \frac \theta 2 f^\dagger_{j\downarrow}, \nonumber\\
   c^\dagger_{j\downarrow} = \sin \frac \theta 2 f_{j\uparrow} + \cos \frac \theta 2 f^\dagger_{j\downarrow},
\end{eqnarray}
where $t+i\Delta=\sqrt{t^2+\Delta^2} e^{-i\theta}$ we can rewrite Eq.~(\ref{eq:H}) in the form
\begin{eqnarray}
  \label{eq:H-bogo}
  H&=&-\sqrt{t^2+\Delta^2}\, \sum_{j,\sigma} (f^\dagger_{j+1\sigma} f_{j\sigma} + f^\dagger_{j\sigma} f_{j+1\sigma}) \nonumber \\
  & &+U\sum_j  \left(f^\dagger_{j\uparrow} f_{j\uparrow}-\frac 1 2\right) \left(f^\dagger_{j\downarrow} f_{j\downarrow} -\frac 1 2\right) \nonumber \\
  & &-\frac{\mu t}{\sqrt{t^2+\Delta^2}} \sum_j (f^\dagger_{j\uparrow} f_{j\uparrow}+f^\dagger_{j\downarrow} f_{j\downarrow}) \nonumber \\
 & &-\frac{\mu \Delta}{\sqrt{t^2+\Delta^2}}  \sum_j (f^\dagger_{j\uparrow} f^\dagger_{j\downarrow}+f_{j\downarrow} f_{j\uparrow}),  \label{eq:chempot}
\end{eqnarray}
showing that $\mu$ is giving both a chemical potential and an s-wave pairing interaction for  the $f$ fermions.  Eq.~(\ref{eq:H-bogo}) is $\mathrm{SU(2)}$ symmetric, and since the Bogoliubov rotation, Eq.~(\ref{eq:bogoliubov}), leaves the expression of the spin operators in fermions invariant, this rules out all non-$\mathrm{SU(2)}$ singlet ground states for the Hamiltonian of Eq.~(\ref{eq:H}) such as spin-density wave or triplet superconductivity.

\subsection{Half-filling}
\label{subsec:half-filling}

Let us first consider the half-filled case. For $\mu=0$, the Hamiltonian~(\ref{eq:H-bogo}) reduces to:
\begin{eqnarray}
    \label{eq:hamiltonian-bog}
    H&=&- \sqrt{t^2+\Delta^2} \, \sum_{j,r=\pm} (f^\dagger_{j+1r} f_{jr}+f^\dagger_{jr} f_{j+1r}) \nonumber \\ & & +U \sum_j  \left(n_{j+} -\frac 1 2\right)   \left(n_{j-} -\frac 1 2\right)  .
\end{eqnarray}
For $U>0$ the Hamiltonian~(\ref{eq:hamiltonian-bog}) has a charge gap and a $c=1$ gapless spin mode (Mott insulator), while for $U<0$ it has gapped spin modes and a $c=1$ gapless charge mode (Luther-Emery liquid).
Since $n_{j\uparrow}-n_{j\downarrow}=n_{j+}-n_{j-}$ and $c^\dagger_{j\uparrow}c_{j\downarrow}=f^\dagger_{j+} f_{j-}$, the spin density wave correlations of the original $c$ fermions are the same as the ones of the $f$ fermions.
However, if we turn to the density, since
\begin{eqnarray}
    \label{eq:density-bogo}
    n_{j\uparrow} + n_{j\downarrow}-1=\cos \theta (n_{j+}+n_{j-} -1) -\sin \theta (f^\dagger_{j+} f^\dagger_{j-} + f_{j-} f_{j+}),
\end{eqnarray}
the density-density correlations of the $c$ fermions are a weighted sum of those of the $f$ fermions and the s-wave superfluid correlations of the same $f$ fermions. In the Mott insulating state, both of those correlations are decaying exponentially, so the density-density correlations show an exponential decay. In the Luther-Emery liquid, the density-density correlations of the $c$ fermions decay as a power law with distance. However, the superfluid component contributes $\sin^2\theta /j^{1/K_c}$ to the density-density correlations, where $K_c$ is the charge Luttinger exponent. Since in the Hubbard model with attractive interaction, $K_c >1$, this contribution dominates the  $\cos^2\theta/j^2$ term coming from the density-density correlations.

To summarize, at half-filling the ground state is always a phase with $c=1$:  A Mott insulator when $U>0$,  a Luther-Emery liquid when $U<0$. But, the Bogoliubov rotation~(\ref{eq:bogoliubov}), turns the density-density correlations and the s-wave superfluid correlation functions into weighted sums of the same correlation functions in the standard 1D Hubbard model. Now, we turn to bosonization \cite{Giamarchi} to consider the effect of the chemical potential $\mu$.

\subsection{Bosonization approach}
\label{sec:chem}

With  $\mu\ne 0$,  we need to consider Eq.~(\ref{eq:chempot}).
The third term of Eq.~(\ref{eq:chempot}) is a chemical potential for the $f$ fermions, while the last term is an s-wave pairing.
To make further progress, it is necessary to use the bosonized representation~\cite{Giamarchi} of the Hamitonian (\ref{eq:chempot}). We find
\begin{eqnarray}
    \label{eq:bosonized}
    H&=&H_c+H_s+H_{cs}, \\
    \label{eq:charge}
    H_c&=&\int \frac{dx}{2\pi} \left[u_c K_c (\pi \Pi_c)^2 + \frac{u_c}{K_c} (\partial_x \phi_c)^2\right] + \frac{\sqrt{2} h}{\pi} \int dx \partial_x \phi_c \nonumber\\& & -\frac{2g_3}{(2\pi \alpha)^2} \int dx \cos \sqrt{8} \phi_c, \\
    \label{eq:spin}
    H_s&=&\int \frac{dx}{2\pi} \left[u_s K_s (\pi \Pi_s)^2 + \frac{u_s}{K_s} (\partial_x \phi_s)^2\right] +\frac{2g_{1\perp}}{(2\pi \alpha)^2} \int dx \cos \sqrt{8} \phi_s, \\
    \label{eq:spincharge}
    H_{cs}&=&-\frac{2\Omega}{\pi \alpha} \int dx \cos \sqrt{2} \theta_c \cos \sqrt{2} \phi_s,
\end{eqnarray}
where $[\phi_\nu(x),\Pi_{\nu'}(x')]=i\delta(x-x') \delta_{\nu,\nu'}$ and $\pi \Pi_\nu=\partial_x \theta_\nu$, with $\nu=c$ for charge excitations and $\nu=s$ for spin excitations.
The short distance cutoff, of the order of the lattice spacing is $\alpha$, and the parameters in the bosonized Hamiltonian are given by
\begin{eqnarray}
    \label{eq:parameters}
    u_c K_c =u_s K_s = v_F, \nonumber \\
    \frac{u_c}{K_c}= v_F + \frac{U \alpha}{\pi}, \nonumber  \\
    \frac{u_s}{K_s}= v_F - \frac{U \alpha}{\pi}, \nonumber  \\
    g_3  = g_{1\perp} = U \alpha, \nonumber \\
    h=\frac{\mu t}{\sqrt{t^2+\Delta^2}}, \nonumber \\
    \Omega=\frac{\mu \Delta}{\sqrt{t^2+\Delta^2}},
\end{eqnarray}
with $v_F=2\sqrt{t^2+\Delta^2}\alpha$.
The simplest case is $U<0$. The spin Hamiltonian $H_s$ is gapped with $\langle \phi_s\rangle=0$, while the charge Hamiltonian $H_c$ remains gapless. Physically, the fermions are already paired, but there is only quasi-long range superconducting order. With $\Omega \ne 0$, the pairing term $\sim e^{i\sqrt{2} \theta_c}$  simply provides the necessary  symmetry breaking and gives rise to a long range superconducting order with $c=0$.

The case of $U>0$ is more complicated. The Hamiltonian $H_c$ is gapped, with $\langle \phi_c\rangle =0$, while $H_s$ is gapless. However, the term proportional to $\Omega$ gives rise to a fully gapped ground state with $\langle \theta_c\rangle=0$ and $\langle \phi_s \rangle =0$. Since the dual fields $\theta_c$ and $\phi_c$ cannot be ordered simultaneously, the $g_3$ and $\Omega$ term are competing with each other and a phase transition is expected.  There are two different scenarios for the transition depending on the ratio $\Delta/t$.
First, in the absence of the chemical potential, the transition is driven by $\Omega$. A weak chemical potential term ($t\ll \Delta$), simply drives the charge doping in the superconducting phase.  A more detailed picture of that scenario is discussed in the Appendix.
Second, in the absence of charge-spin coupling~(\ref{eq:spincharge}), the $h$ term would give rise to a commensurate-incommensurate transition \cite{Giamarchi,japaridze_cic_transition,pokrovsky_talapov_prl} closing the charge gap, yielding  a two-component Tomonaga-Luttinger liquid having two $c=1$ modes for spin and charge of different velocities. At the transition point \cite{Giamarchi}, we would have $K_c=1/2$ with $K_s=1$ because of $\mathrm{SU(2)}$ spin symmetry. Now,  Eq.~(\ref{eq:spincharge}), implies  that the scaling dimension of $\Omega$  at the transition point  is $2-1/(2K_c)-K_s/2=1/2$, so that for $0<|\Delta| \ll t$,  a gap $M \sim \frac{v_F} \alpha \left(\frac{\Omega \alpha}{v_F}\right)^2$ immediately opens both in the spin and the charge modes. The superconducting state with $\langle \theta_c\rangle=0$ and $\langle \phi_s\rangle=0$ is then formed.

These two scenarios can be distinguished by considering the evolution of the charge density with $\mu$. In both cases, there is according to Eq.~(\ref{eq:density-bogo}) a  contribution proportional to $\langle \cos \sqrt{2} \theta_c \cos \sqrt{2} \phi_s\rangle $. That contribution is non-singular since the charge modes are always gapped, and can be obtained from linear response.  However, in the vicinity of the commensurate-incommensurate transition \cite{japaridze_cic_transition,pokrovsky_talapov_prl}, the particle density $\langle n_++n_- -1 \rangle \sim (h-h_c)^{1/2}$ giving rise to a kink in the density at the transition between the  Mott insulator and the superfluid. Summing the two contributions, the fermion density varies as
\begin{eqnarray}
    \langle n_{j\uparrow} + n_{j\downarrow} -1 \rangle = A(\Delta,t,U) \mu + B(\Delta,t,U) \sqrt{\mu -\mu_c}\Theta(\mu-\mu_c),
\end{eqnarray}
when $\Delta \ll t$. By contrast, when $\Delta \gg t$,  a large gap is present on both sides of the transition,  and the density varies smoothly with the chemical potential.
\begin{eqnarray}
    \langle n_{j\uparrow} + n_{j\downarrow} -1 \rangle = C(\Delta,t,U) \mu + D(\Delta,t,U) f(\mu),
\end{eqnarray}
In the Appendix, we show that at the transition, $f(\mu) \sim (\mu -\mu_c) \ln |\mu-\mu_c|$, \textit{i.e.} there is only a vertical tangent instead of a slope discontinuity.

\subsection{Friedel oscillations}
\label{sec:friedel}

So far, we have considered an infinite chain. With a finite chain of $N$ sites,
we have to introduce two fictitious sites $0$ and $N+1$ such that
\begin{equation}
  \label{eq:obc-fermion}
  c_{0\sigma}=c_{N+1\sigma}=0.
\end{equation}
Using Eq.~(\ref{eq:bogoliubov}) this translates into  $f_{0+}=f^\dagger_{0-}=0$ and  $f_{N+1+}=f^\dagger_{N+1-}=0$. Thus, the $f$ fermions obey  the same open boundary conditions as the original fermions. In bosonization, the boson fields \cite{Giamarchi,Fabrizio1995} in Eqs.~(\ref{eq:charge})--(\ref{eq:spin}) have to satisfy
\begin{equation}
  \label{eq:obc-boson}
  \phi_\nu(0)=\phi_\nu(N+1)=0,
\end{equation}
for $\nu=c,s$.
In the MI phase, the conditions~(\ref{eq:obc-boson}) are already satisfied in the bulk  by the charge modes. As the edge does not perturb the ordering of the charge modes, only its effect the spin modes needs to be analyzed. In facts \cite{eggert_openchains,eggert_openchains_short,rommer00_friedel1d},  the boundary conditions being $\mathrm{SU(2)}$ symmetric, only $\langle \cos \sqrt{2} \phi_s \rangle(j)  \sim j^{-1/2} \ne 0$ (near the left edge). Taking into account the ordering of the charge modes, only the Bond Order Wave (BOW) order parameter $\langle \cos \sqrt{2} \phi_c \cos \sqrt{2} \phi_s \rangle \sim j^{-1/2}$ expectation value shows power law oscillations. This translates into
\begin{eqnarray}\label{eq:friedel-bow}
\langle \sum_\sigma (c^\dagger_{j+1\sigma} c_{j\sigma} + c^\dagger_{j\sigma} c_{j+1\sigma}) \rangle \sim \frac{(-)^j}{\sqrt{\frac{N+1} \pi \sin \left(\frac{\pi j}{N+1}\right)}}.
\end{eqnarray}

In the SC phase, the conditions~(\ref{eq:obc-boson}) are already satisfied in the bulk by the spin modes. For the charge modes, however, the boundary conditions~(\ref{eq:obc-boson}) impose that the superconducting order parameter vanishes at the edge and make the  BOW order parameter  nonzero.  Moving into the bulk, the BOW operator expectation value decays exponentially, while the SC order parameter recovers its bulk expectation value.

\section{Numerical methods}

\subsection{MPO formalism}

We shall now focus exclusively on the (more complicated) $U>0$ case and use both finite and infinite size DMRG. Since the Hamiltonian has  translation symmetry, we construct
the corresponding matrix product operator (MPO), which acts on matrix product state
(MPS). By performing the usual matrix multiplication, we can check
that the following MPO does represent our Hamiltonian:
\begin{equation}
\left( \begin{array}{cccccc} 1 & c^{\dagger}_{i\uparrow} &
c^{\dagger}_{i\downarrow} & c_{i\uparrow} & c_{i\downarrow} &
U(n_{i\uparrow}-\frac{1}{2})(n_{i\downarrow}-\frac{1}{2})
-\mu (n_{i\uparrow}+n_{i\downarrow}) \\
0 & 0 & 0 & 0 & 0 & -t c_{i\uparrow}   - \Delta c^{\dagger}_{i\downarrow} \\
0 & 0 & 0 & 0 & 0 & -t c_{i\downarrow} + \Delta c^{\dagger}_{i\uparrow} \\
0 & 0 & 0 & 0 & 0 &  t c^{\dagger}_{i\uparrow}   + \Delta c_{i\downarrow} \\
0 & 0 & 0 & 0 & 0 &  t c^{\dagger}_{i\downarrow} - \Delta c_{i\uparrow} \\
0 & 0 & 0 & 0 & 0 & 1
\end{array} \right)
\end{equation}
where we omit the boundary operators. Obviously, we need to take care
of ordering for
fermions when we carry out iDMRG by acting with the MPO on the MPS.

Let us assume that the physical index $\sigma_{i}$ labels the
state on the $i$-th site. For the Hubbard model,
$\sigma_{i}=(\alpha_{i},\beta_{i})$, where $\alpha_{i} (\beta_{i}) = 0$ or $1$ means that there is a
vacancy or occupation of the spin-up (down) fermion at the $i$-th
site, respectively. The state of the Fock space
for a $L$-lattice system is thus written in terms of the creation
operators $c^{\dagger}_{i\uparrow}$ and
$c^{\dagger}_{i\downarrow}$ as follows:
\begin{equation}
| \sigma_{0} \cdots \sigma_{L-1} \rangle =
(c^{\dagger}_{0\uparrow})^{\alpha_{0}}
(c^{\dagger}_{0\downarrow})^{\beta_{0}} \cdots
(c^{\dagger}_{L-1\uparrow})^{\alpha_{L-1}}
(c^{\dagger}_{L-1\downarrow})^{\beta_{L-1}}  |0 \rangle \, .
\end{equation}
It is important to maintain the order of the
fermions in the state of the Fock space to handle the minus
sign caused by the exchange of fermion. We adopt the order of
spin-up first and spin-down next as above.

For iDMRG with a two-site unit cell, two tensors, $A$ and $B$, in
the MPS are repeated as $\cdots ABABAB \cdots$ with the usual
matrix multiplication. The tensors, $A_{ab}^{\sigma}$ and
$B_{cd}^{\rho}$, have three indices, among which the physical
index $\sigma$ and $\rho$ takes a value from $0$ to $3$ for our
model. For the degree of freedom of the internal bond, the indices
$a$ (left) and $b$ (right) for $A$ range from $0$ to $\chi-1$,
where $\chi$ is the dimension of the internal bond. The Schmidt
coefficients between $A$ and $B$, and between $B$ and $A$, are
denoted by $\lambda^{AB}$ and $\lambda^{BA}$, respectively. Thus,
a state in the space of matrix product states is written as
\begin{equation}
|\Psi \rangle = \sum_{\cdots \sigma\rho\nu\eta\cdots}
\mbox{Tr}(\cdots A^{\sigma}_{ab} \lambda^{AB}_{b} B^{\rho}_{bc}
\lambda^{BA}_{c} A^{\nu}_{cd} \lambda^{AB}_{d} B^{\eta}_{de}
\cdots )| \cdots \sigma\rho\nu\eta\cdots \rangle  ,   \label{eq:MPS}
\end{equation}
where $\mbox{Tr}$ means that the indices of the internal bonds $a,
b, c, d, \cdots$ are summed up. Scaling properties will be sought
by increasing $\chi$

Regardless of the $t$, $U$, $\mu$, and $\Delta$ values used in our
calculations, we have observed a smooth convergence. Our numerical
DMRG results show that the ground-state solutions fall into two
classes: MPS are either of the form $\cdots ABABAB\cdots$ (i.e.
unit cell of two sites that we will identify as a Mott phase) near
half filling ($\mu=0$), or uniform $\cdots AAAAAA\cdots$ further
away from half-filling, that we will identify as metallic or
superconducting phases, for $\Delta=0$ or non-zero, respectively.

\subsection{Phase diagram}

We will present data obtained using the infinite DMRG (with a two-site unit cell) as well as the finite-size algorithm for chain length up  to $L=512$.
After computing the ground state $| \Psi \rangle$, we compute local quantities such as the bond energy and the local densities and we also use the half-chain entanglement entropy to determine if the system is critical and, if so, what is its central charge.

By contraction of the Hamiltonian bond operator with the ground-state MPS, we obtain the
bond energy.
Close to half-filling, since the MPS has an ABAB form,   we
obtain alternating bond energies on even and odd bonds. We have observed however that the modulation seems to vanish for $\chi\rightarrow\infty$.
To be more quantitative, we determine the half-chain entanglement entropy $S$, which
is related to the Schmidt coefficients $\lambda_{a}$  as
\begin{equation}
S = -\sum^{\chi - 1}_{a=0}\lambda^{2}_{a} \ln
\lambda^{2}_{a}. \label{eq:EE2}
\end{equation}
The Schmidt coefficients $\lambda_{a}$ are obtained when we
perform a singular value decomposition (SVD) to find the matrices
$A$ and $B$ in the MPS. Normalization of $\sum^{\chi -
1}_{a=0}\lambda^{2}_{a}=1$ guarantees $\langle \Psi | \Psi \rangle
= 1$. For the $\cdots ABABAB \cdots$ solution, we obtain two
different values of the entanglement entropy: $S_{o}$ with
$\lambda^{AB}$ on the odd bonds and $S_{e}$ with $\lambda^{BA}$ on
the even ones.

\begin{figure*}
\includegraphics[width= 12.0 cm]{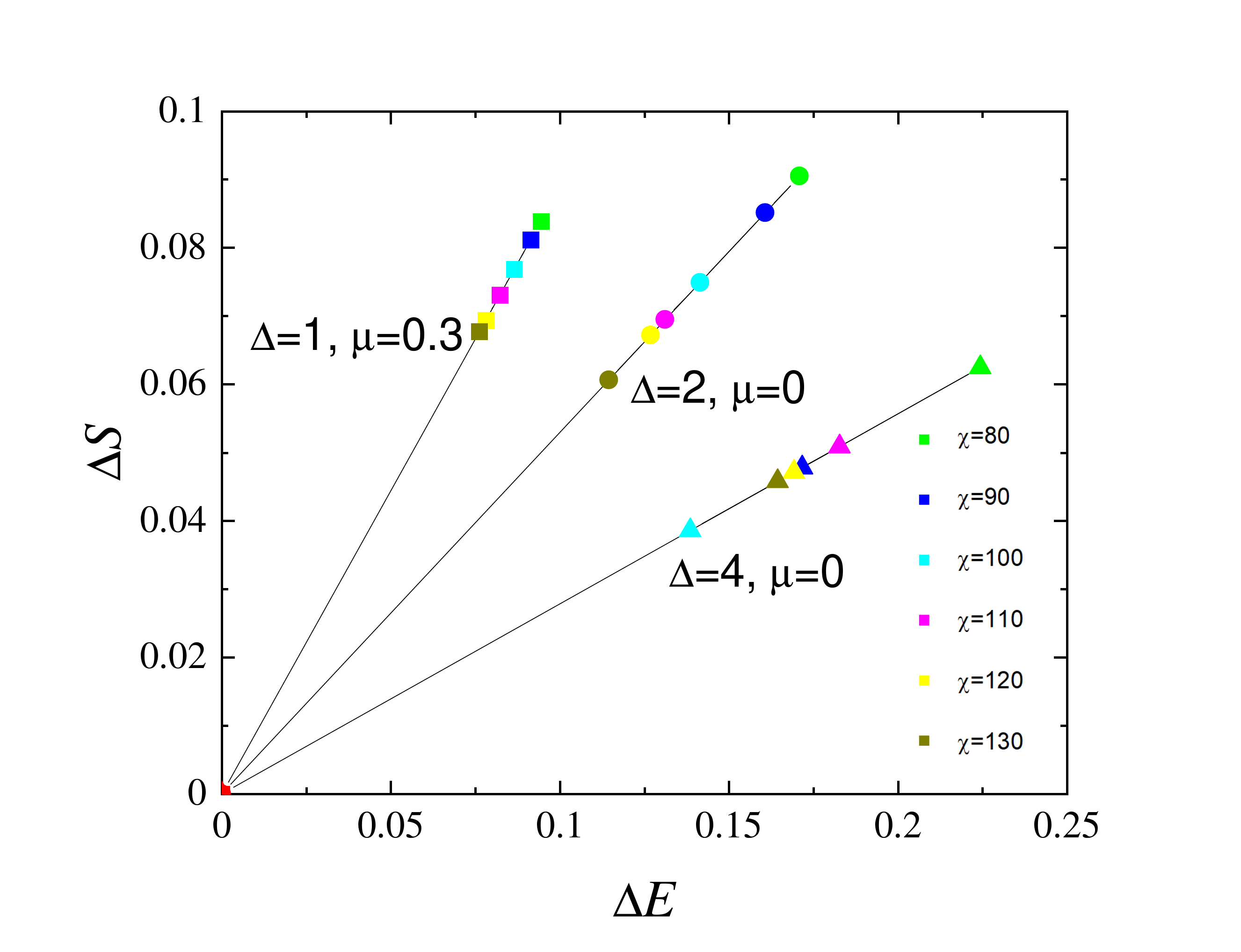}
\caption {The entanglement entropy difference $\Delta S$ versus
the bond energy difference $\Delta E$ for three cases by setting
$t=1$ and $U=4$. The different colors of the data points are
corresponding to the difference values of $\chi$. We note the
particular order of the data points for the case of $\Delta=4$,
$\mu=0$. The linear fits are compatible with a zero intercept,
i.e. $\Delta S$ and $\Delta E$ are proportional.}
\label{fig:Delta}
\end{figure*}

The calculation shows that, close to half-filling, both the bond
energy and the entanglement entropy have a finite modulation. In
such a case,  the energy difference $\Delta E = E_{e} - E_{o}$ and
the entropy difference $\Delta S =S_{o} - S_{e}$ are proportional
to each other~\cite{Laflorencie}. In Fig.~\ref{fig:Delta}, we
present the $\chi$-dependence of $\Delta E$ versus $\Delta S$. The
numerical results confirm that $\Delta S$ is proportional to
$\Delta E$ to a very high accuracy of $10^{-6}$. We conclude that
$\Delta S = 0$ and $\Delta E = 0$ for the infinite bond dimension
of $\chi = \infty$, as expected.

\begin{figure*}
\includegraphics[width= 12.0cm]{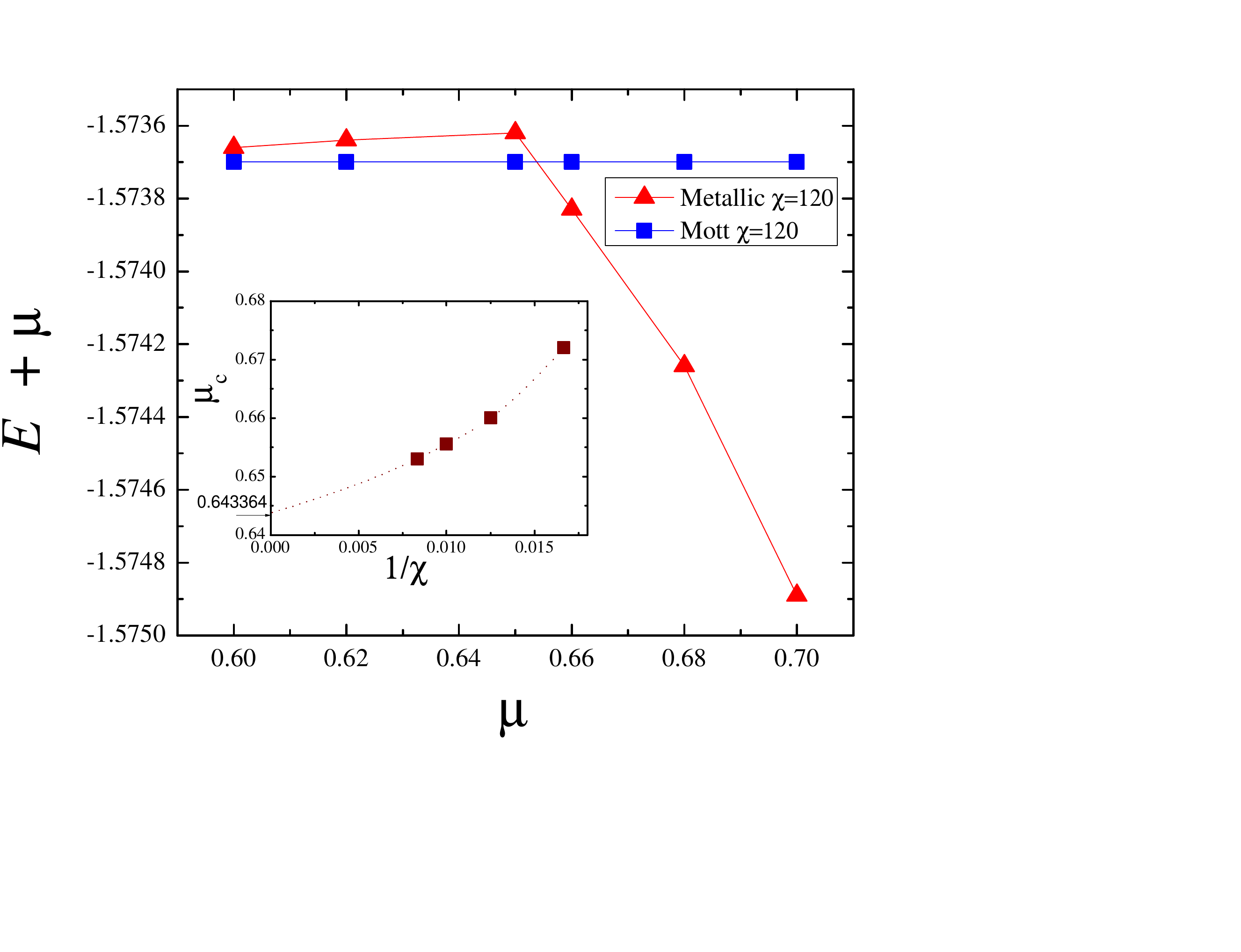}
\caption {Ground state energy (shifted by $\mu$) versus $\mu$
showing a level crossing between the Mott-insulating and the
metallic iDMRG solutions, at a finite $\chi$ value. We have set
$t=1$, $U=4$ and $\Delta=0$. Inset: the scaling of the crossing
points with $1/\chi$ agrees with the exact transition point at
$\mu_{+} \simeq 0.643364$.} \label{fig:crossing}
\end{figure*}

In order to determine the Mott transition, characterized by a
change in the compressibility, we compute the (average) ground
state energy $E = (E_{e} + E_{o})/2$ vs the chemical potential
$\mu$ starting from $\mu=0$. On the other hand, we can also
compute the ground-state energy in the uniform solution by
decreasing $\mu$ (starting from large values). In each iDMRG
calculation, the tensors of the initial environment are given by
the previous solution of the different $\mu$. First, as a
benchmark, we plot in Fig. \ref{fig:crossing} the evolution of the
ground state energy for $\Delta=0$, where we find a level-crossing
at a critical $\mu_c$. In full care of the entanglement,
corresponding to $\chi \rightarrow \infty$, our extrapolation of
$\mu_c$ is quite close to the exact value $\mu_+$ found using the
Lieb-Wu method~\cite{Yang,hubbard_book} and it corresponds to the
well-known second-order phase transition between an incompressible
Mott phase and a metallic one.

\begin{figure*}
\includegraphics[width= 12.0cm]{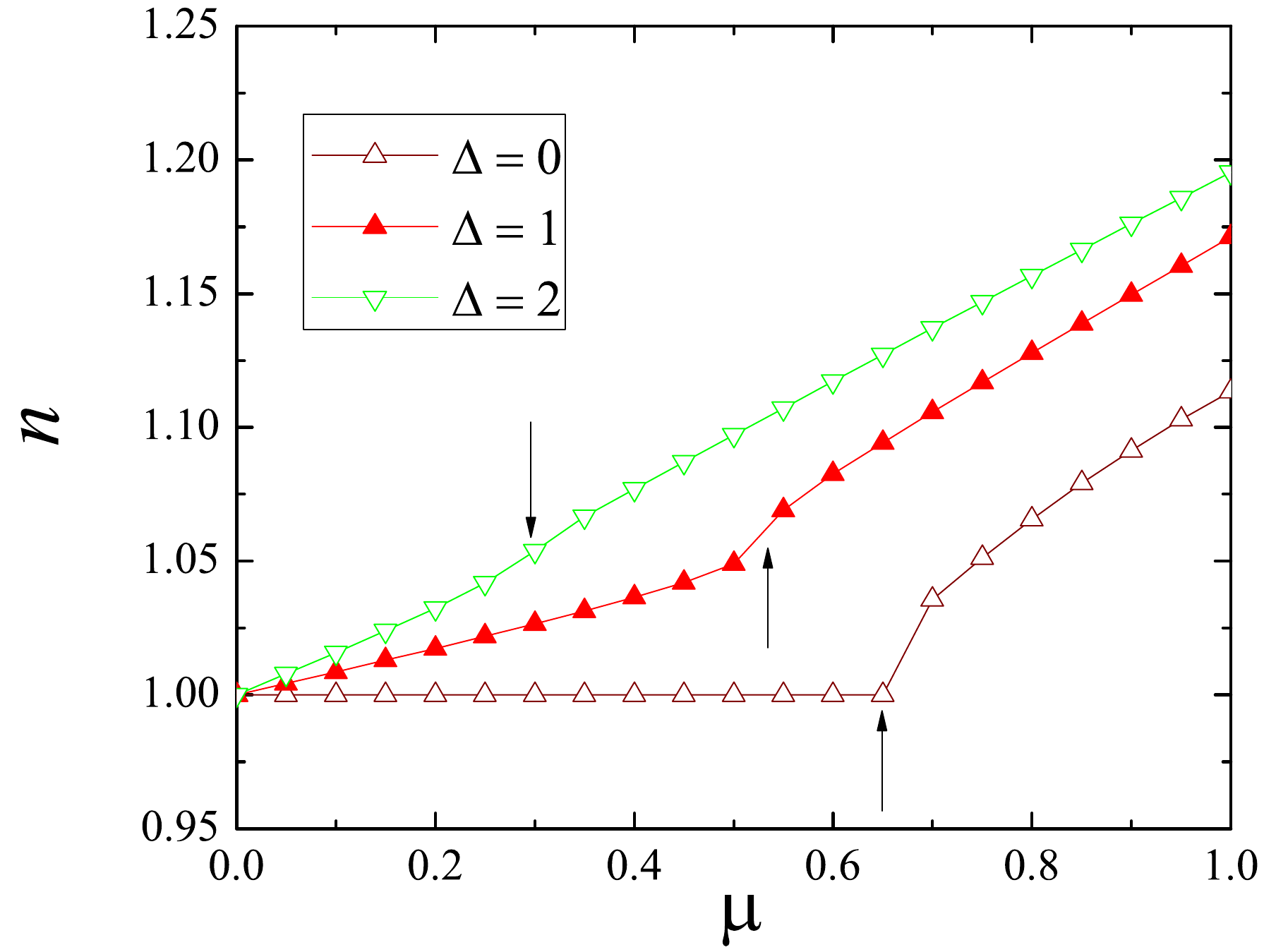}
\caption {Density $n$ versus chemical potential $\mu$ for
$U=4$, $t=1$ and $\chi=120$. For different values
$\Delta$, we observe an abrupt change of the slope $\partial n/\partial\mu$.
} \label{fig:num}
\end{figure*}

We also compute the density, i.e. the expectation value of the
occupation number $n=\langle
n_{i,\uparrow}+n_{i,\downarrow}\rangle$, as a function of $\mu$ as
shown in Fig. \ref{fig:num} for several values of the pairing
strength $\Delta$ and $U=4$. For $\Delta=0$, we do observe an
incompressible phase around $\mu=0$ and a transition point
identical to the previous one, see Fig.~\ref{fig:crossing}. For
$\Delta>0$, the compressibility (which is the slope of $n$ vs
$\mu$) is always finite but we do observe a sudden change for some
critical $\mu_c$, which we identify as the phase transition
between Mott and superconducting phases. The finite compressibility is a consequence of Eq.~(\ref{eq:bogoliubov}). The fermion density is a weighted sum of the density of the Bogoliubov quasiparticles and of their superconducting order parameter. The latter responds linearly to $\mu$ according to Eq.~(\ref{eq:H-bogo}) resulting in a nonzero compressibility even in the Mott phase.

\begin{figure*}[!ht]
\includegraphics[width= 16.0 cm]{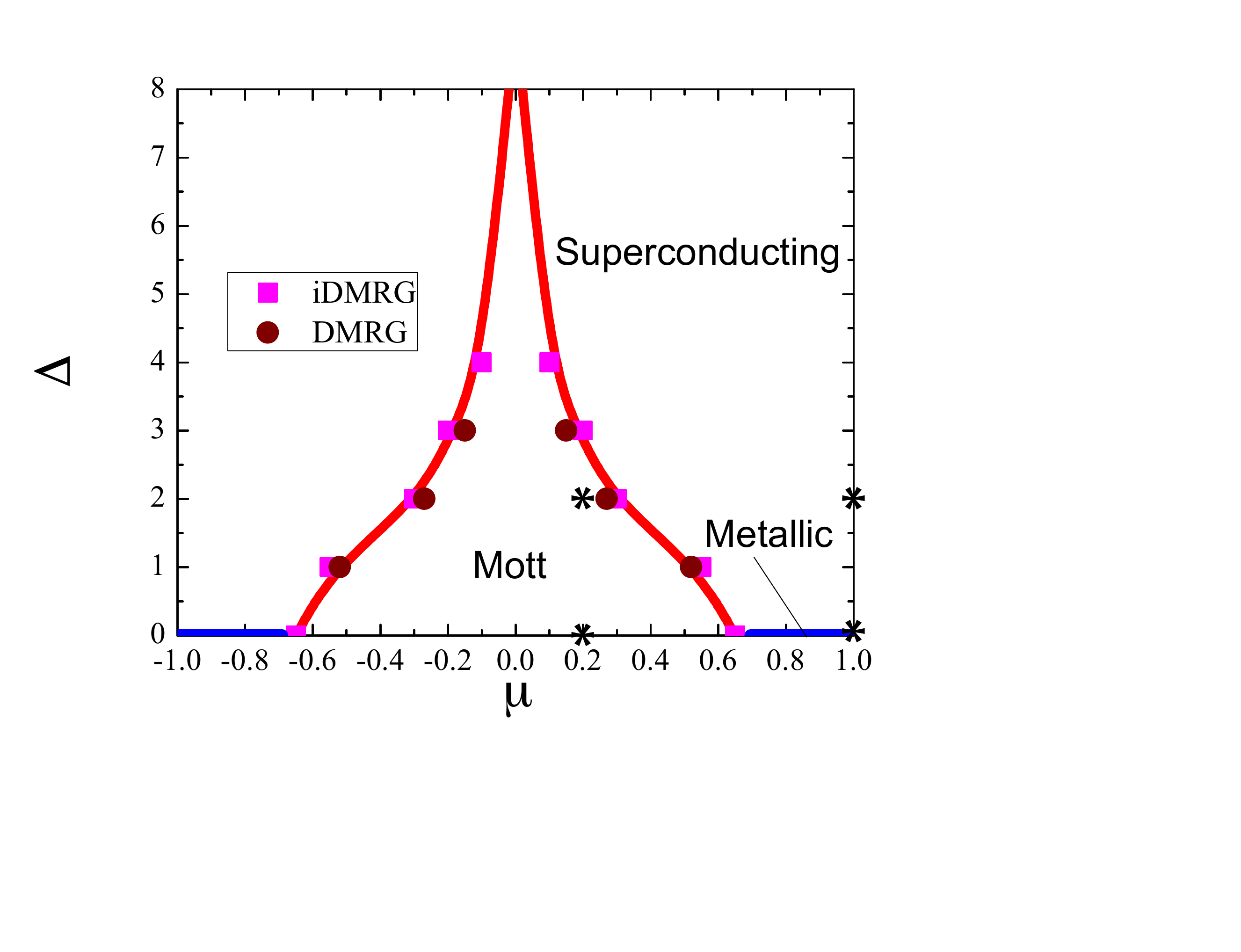}
\caption {Numerical phase diagram of the one-dimensional Hubbard
model at $t=1$ and $U=4$, as a function of the chemical potential
$\mu$ and the singlet pairing potential $\Delta$, obtained from
iDMRG (magenta squares) and DMRG (purple circles, $L=128$). The
red line is a guide to the eyes. The region of the Mott-insulating
phase shrinks for smaller $U$. Star symbols correspond to the points where local quantities are plotted in Fig.~\ref{fig:DMRG_local}.} \label{fig:Dia}
\end{figure*}

As a concluding remark about this section, we have observed that the critical $\mu_c$ varies with the pairing strength $\Delta$ so that we can summarize the numerical results  in the phase diagram shown in Fig.~\ref{fig:Dia}. On top of our numerical data, we provide a qualitative sketch of the full phase diagram but it is difficult numerically to determine what happens for large $\Delta$ at $\mu=0$.  In this region, we can use the partial particle-hole transformation that was discussed in Sec.~\ref{sec:model}. Indeed, for $\mu=0$, the model with parameters $(t=1,U,\Delta)$ at large $\Delta$, which is difficult to analyze, is equivalent to the one at $(t=\Delta,U,\Delta=1)$ which is simply a tight-binding chain with small perturbation. In this case, we do expect a Mott phase with a very small gap~\cite{hubbard_book}, hence a very small Mott region.

\subsection{Local observables and exponents}
\begin{figure*}[!ht]
\includegraphics[width= 16.0 cm]{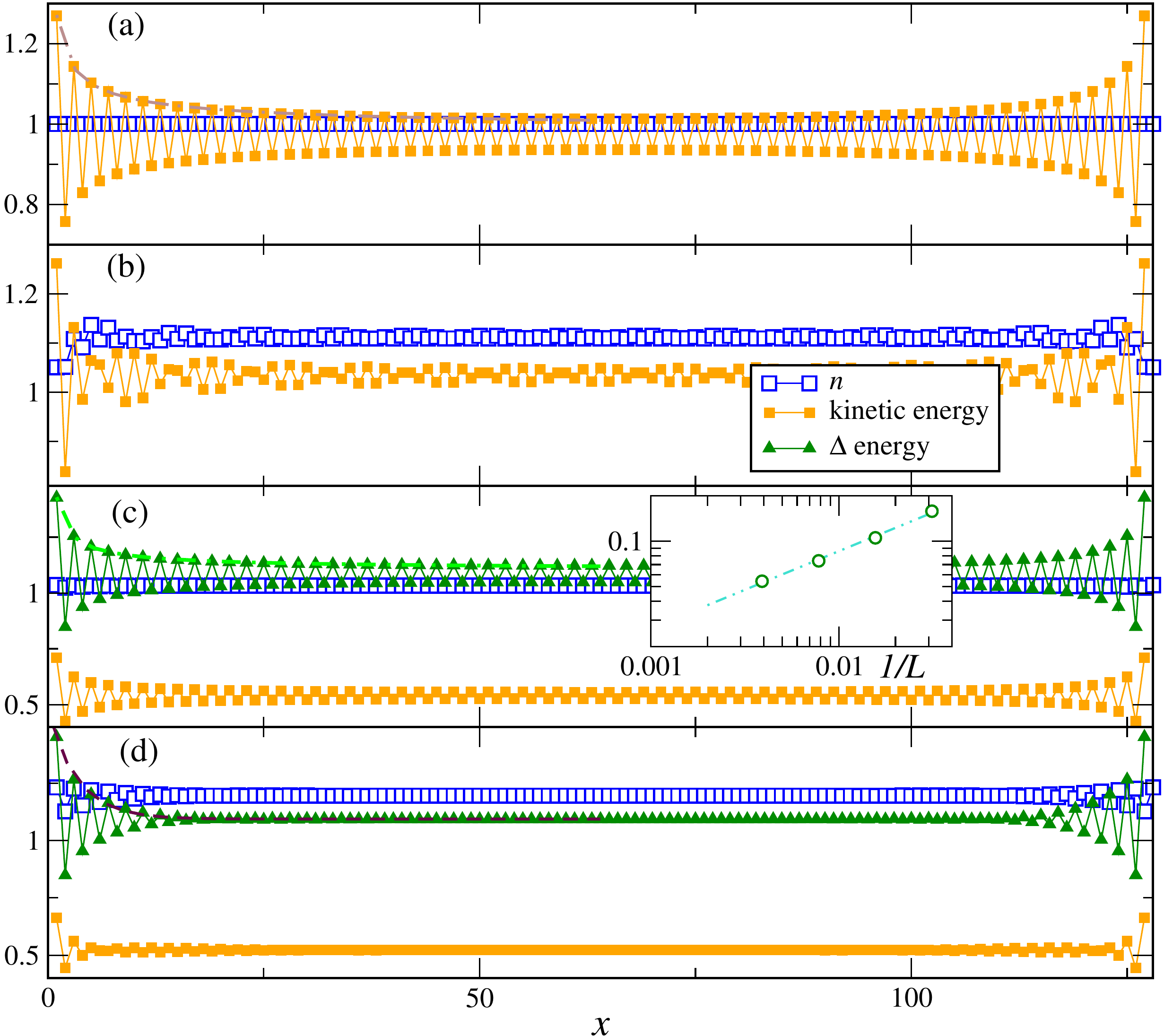}
\caption {Local quantities of the one-dimensional Hubbard
  model at $t=1$ and $U=4$, obtained from DMRG on chains of length $L=128$ at various points in the phase diagram (star symbols in Fig.~\ref{fig:Dia}). From top to bottom, parameters $(\mu,\Delta)$ are: (a) (0.2,0) in the Mott phase of the pure Hubbard model; (b) (1,0) in the metallic phase of the pure Hubbard model; (c) (0.2,2) in the Mott phase of our model; (d) (1,2) in the superconducting phase. Inset of panel (c) shows the pairing modulation measured in the center of the chain as a function of $1/L$ in logarithmic scale, fitted with an exponent $1/L^{0.51}$.} \label{fig:DMRG_local}
\end{figure*}

In order to confirm that there is no breaking of the translation symmetry in the thermodynamic limit, we also plot in Fig.~\ref{fig:DMRG_local} several local quantities measured by finite-size DMRG (with open boundary conditions) in several points of the phase diagram. Panels (a-b) correspond to the pure Hubbard model respectively in the Mott and metallic phase. In the Mott phase, the density is locked to $n_\uparrow=n_\downarrow=n/2=0.5$ per site since the chemical potential is smaller than the gap and we do observe Friedel oscillations in the bond kinetic energy, similar to the well-known oscillations in Heisenberg spin chains~\cite{Laflorencie}, which can be fitted as $1/x^{0.55}$ in good agreement with the predicted $1/\sqrt{x}$ knowing that logarithmic corrections are present~\cite{Giamarchi}. In the metallic phase for a generic filling, oscillations are incommensurate and can provide accurate information about Luttinger exponents etc~\cite{Fabrizio1995,Friedel_1998}. For finite $\Delta$ and small $|\mu|$, see Fig.~\ref{fig:DMRG_local}c, this is the generalized Mott phase with very small density fluctuations (non-zero since there is a finite compressibility) and power-law Friedel oscillations both in the bond kinetic energy and in the bond pairing energy: we have attempted to fit the pairing modulations with a power-law from the edge, which leads to an exponent 0.71; however, fitting the modulation measured in the center of the chain as a function of $1/L$ leads to an exponent $0.51$ in perfect agreement with analytic prediction, see Eq.~(\ref{eq:friedel-bow}).  Last, in the superconducting phase, see  Fig.~\ref{fig:DMRG_local}d, the Friedel oscillations in all local quantities are short-range and can be fitted with an exponential form, as expected for a fully gapped phase.

Note that the physics of our model is rather different from the extended Hubbard model which is known to host long-range ordered BOW~\cite{Sengupta2002}.

\subsection{Entanglement entropy scaling}

\begin{figure*}
\includegraphics[width= 12.0 cm]{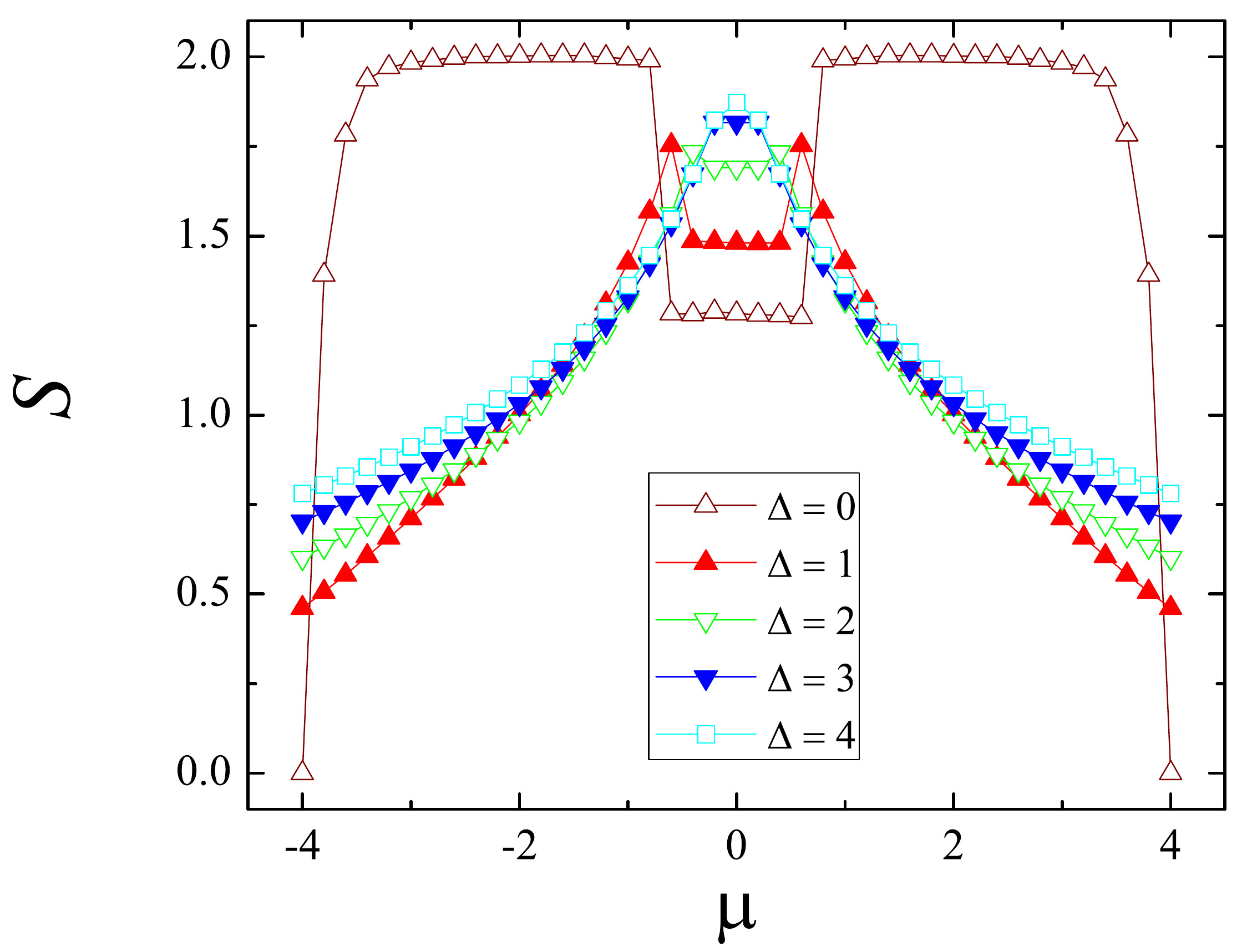}
\caption {Half-chain entanglement entropy versus $\mu$ for
several values of the pairing $\Delta$. We have set the parameters to $t=1$, $U=4$
and $\chi=100$.} \label{fig:EE}
\end{figure*}

It is well-established that block entanglement entropy scaling can be used to determine if the ground-state is gapped or critical. In the later case, the central charge of the underlying Conformal Field Theory (CFT) can also be computed~\cite{Cardy}.
In Fig.~\ref{fig:EE}, we present the half-chain entanglement
entropy $S = (S_{o}+S_{e})/2$, which is obtained from
iDMRG. In agreement with our local measurements from the previous section, we do observe a rather flat plateau region around $\mu=0$, at least for $\Delta$ not too large, corresponding to the Mott phase obtained from the compressibility data. Note that the size of the plateau is decreasing with $\Delta$ so that it is still difficult to determine the physics for large $\Delta$ at $\mu=0$.

\begin{figure*}
\includegraphics[width= 12.0 cm]{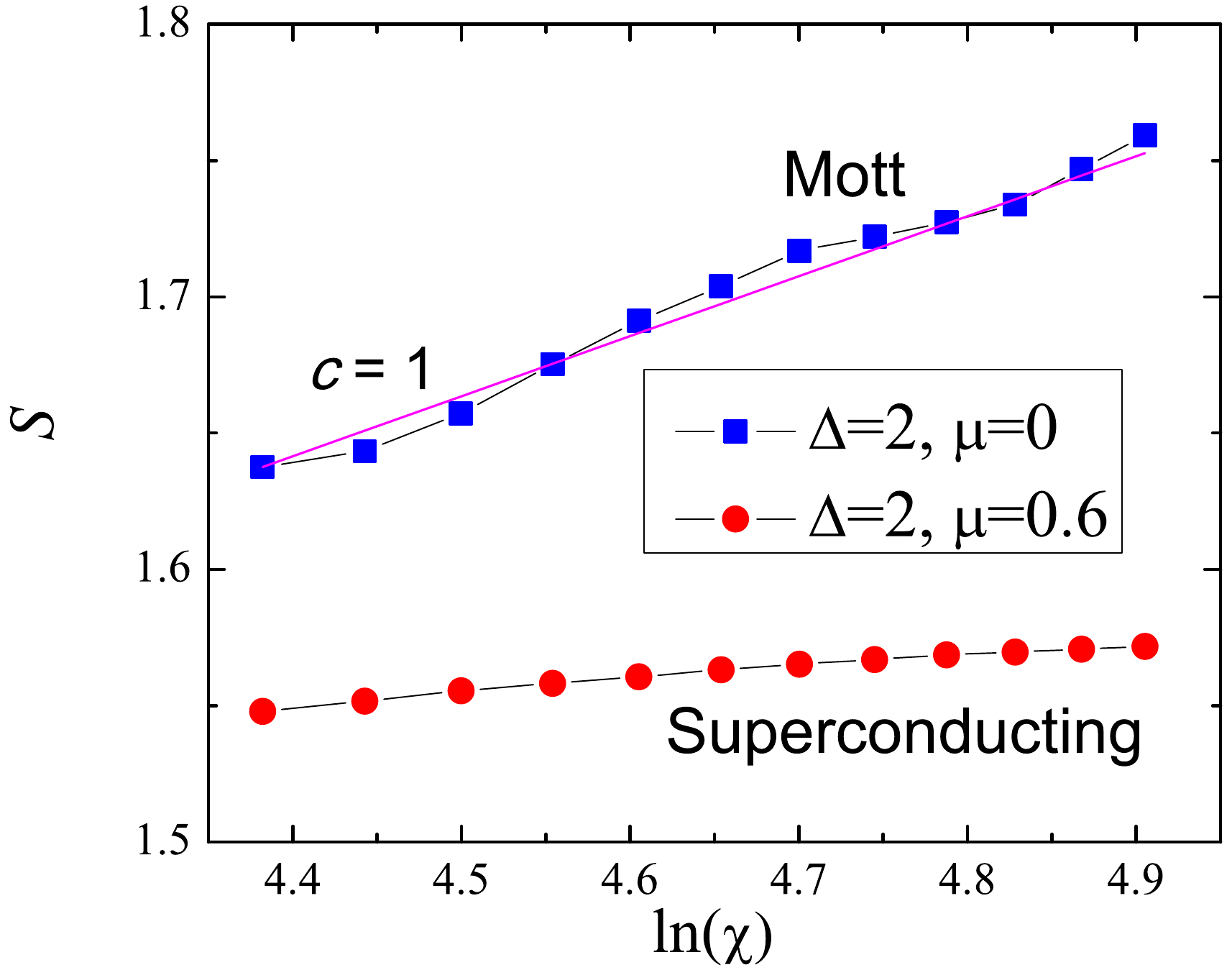}
\caption {Entanglement entropy versus $\ln(\chi)$ for
two values of $\mu$, fixing $U=4$ and $\Delta=2$.
For $\mu=0.6$, $S$ saturates at large $\chi$, in agreement with a gapped superconducting behavior.
At $\mu=0$, the data can be fitted as
$S = 0.22 \ln(\chi)+\tilde{s}$, in agreement with the conformal scaling with central charge $c=1$ (see text).} \label{fig:logchi}
\end{figure*}

In addition, using a conformal scaling with $\chi$,
$$S = \frac{1}{\sqrt{\frac{12}{c}}+1}\ln \chi + \tilde{s},$$
one can determine the central charge $c$~\cite{Zamolodchikov} in
all critical phases~\cite{Pollmann2009} with a constant
$\tilde{s}$. In Fig. \ref{fig:logchi}, we present the
finite-$\chi$ scaling of the half-chain entanglement entropy for
several $\Delta$ and $\mu$. For parameters $U=4$, $\Delta = 2$ and
$\mu = 0.6$, we observe the saturation of $S$ at large $\chi$, a
behavior characteristic of a fully gapped phase as expected for
the superconducting one. In contrast, for the model with $U=4$,
$\Delta=2$ and $\mu=0$, the above conformal scaling is well
realized providing the central charge is set to $c = 1$,  as
expected for a Mott phase with a single gapless spin mode.

In order to provide a complementary quantitative analysis, we have used finite-size DMRG algorithm~\cite{itensor} keeping up to $m=4000$ states and with a discarded weight below $10^{-8}$.
For a finite-system with open boundary conditions, conformal field theory \cite{Cardy} predicts that, in a critical region, the block entanglement
entropy $S$ should follow the universal scaling behavior:
\begin{equation}
S = \frac{c}{6} \ln d(x|L) + \tilde{s} . \label{eq:EEconf}
\end{equation}
where $c$ is the central charge, $d(x|L)=\pi/L \sin(x\pi/L)$ is
the conformal block size of size $x$, and $\tilde{s}$ is a
non-universal constant.

\begin{figure*}
\includegraphics[width= 12.0 cm]{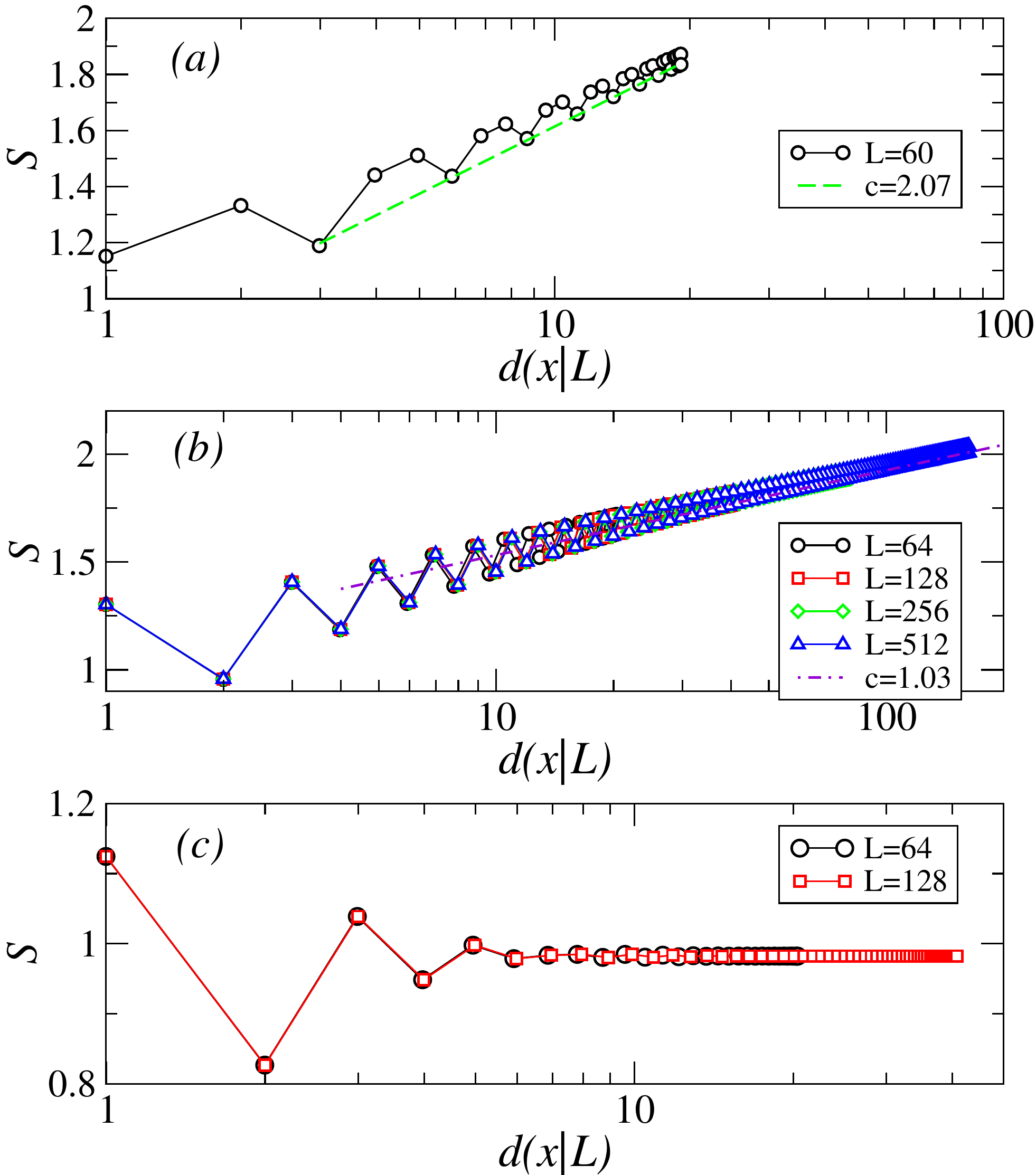}
\caption {Entanglement entropy scaling versus the conformal size of the block
  for $U=4$ and various parameters $\Delta$ and $\mu$, obtained from DMRG on chains of length $L$.
  (a) $\Delta=0$ and $\mu=2$ in the metallic phase;
  (b) $\Delta=2$ and $\mu=0$ in the Mott phase;
(c) $\Delta=2$ and $\mu=2$ in the superconducting phase.}  \label{fig:c}
\end{figure*}

In Fig. \ref{fig:c}, we present the
finite-size scaling of the entanglement entropy for $U=4$ in all different regions. In the metallic phase ($\Delta=0$, $\mu=2$), we measure a large central charge $c\simeq 2$ corresponding to two gapless modes (one in the charge channel, one in the spin channel) as expected. For $\Delta=2$ and $\mu=2$, we are in the fully gapped superconducting phase.
Last, in the Mott phase at or close to half-filling (for instance $\Delta=2$ and $\mu=0$), we observe a smooth crossover from a large central charge $c\simeq 2$ at small distance to a proper $c=1$ at large distance, as expected from a single gapless spin mode, as found for instance in the pure Hubbard model with $\Delta=0$. Indeed, it is well-known that, for the pure Hubbard model at half-filling, the charge gap is exponentially small ($\sim \exp(-t/U)$) while it becomes of order $U$ at large $U$~\cite{hubbard_book}. Similarly, there is a corresponding length scale (proportional to the inverse of the gap) that governs this crossover.

In conclusion on this section, our finite-size DMRG calculations have confirmed that the charge channel is always gapped for finite $\Delta$.  However, we have distinguished the Mott phase from the fully gapped superconducting one by its gapless $c=1$ spin mode. Also, we have not seen any clear  sign of an intermediate BOW phase with spontaneous translation symmetry breaking, which in principle is allowed as discussed in the Appendix.

\section{Conclusion}

In summary, we have used both the finite-size and infinite DMRG to
obtain the ground-state of the one-dimensional
Hubbard model with an additional singlet pairing potential. Such a model would be relevant for a strongly correlated chain with some proximity coupling to a singlet superconductor. We have computed local quantities as well as entanglement properties in order to establish the full phase diagram, including Mott, metallic and superconducting phases.

Our study has revealed a particularly interesting feature of the
Mott and superconducting phases, connected to the existence of a
potential breaking explicitly particle number conservation. In
that case, the inverse of the compressibility $\kappa$ is {\it no
longer} related to the many-body charge gap $\Delta_C$, as
$\kappa^{-1}\sim L\, \Delta_C $, so that $\Delta_C$ and $\kappa$
could be simultaneously non zero in the thermodynamic limit
$L\rightarrow\infty$. Such a remarkable feature is examplified by
the Mott and the superconducting phases which are both
simultaneously gapped (in the charge sector) and compressible. The
Mott phase can however be characterized by the existence of a
gapless spin mode (described by a $c=1$ CFT) while the
superconducting phase is fully gapped.

It would be an interesting prospect to extend
this study to two-dimensional systems, using for instance PEPS formulation that does not suffer from the negative sign problem.

\section*{Acknowledgments}
This work was partially supported by the Basic Science Research
Program through the National Research Foundation of Korea (NRF)
funded by the Ministry of Education, Science and Technology (Grant
No. NRF-2017R1D1A1A0201845 to M.H.C.). The authors would like to thank J.Y. Chen for helpful discussions.
DP acknowledges support by the TNSTRONG  ANR-16-CE30-0025  and TNTOP ANR-18-CE30-0026-01 grants awarded by the French  Research  Council.
This work was granted access to the HPC resources of CALMIP supercomputing center under the allocations P1231. M.H.C. appreciates the hospitality of CNRS at Toulouse, where this work was initiated during his sabbatical year.

\section*{Appendix: Mott Insulator to superconductor transition}

In the present appendix, we give a more detailed discussion of the case $\mu\ne 0$ where the $g_3$ and $\Omega$ terms of  (\ref{eq:charge})--(\ref{eq:spincharge}) are computed. A similar competition  was discussed in the case of the ionic Hubbard model~\cite{fabrizio_dsg,tincani} where three phases were found, a Mott insulator, a Band insulator and a narrow~\cite{tincani} intermediate bond order wave phase. Analogously, in our model, an intermediate $c=0$ bond order wave (BOW) phase with $\langle \phi_c \rangle=0$ and $\langle \phi_s\rangle=0$ could exist between the Mott insulator (MI) and the superconductor (SC). The MI-BOW transition is a Berezinskii-Kosterlitz-Thouless transition where a gap in the spin modes opens, leading to $\langle \phi_s \rangle=0$. The BOW-SC transition only affects the charge sector. Thus, to discuss that transition, we can replace $\cos \sqrt{2} \phi_s$ by its expectation value in Eq.~(\ref{eq:spincharge}).
At the special point $K_c=1/2$, using the rescaling $\phi=\sqrt{2} \phi_c$ and $\theta = \theta_c/\sqrt{2}$ we can rewrite the low-energy Hamiltonian~(\ref{eq:charge})--(\ref{eq:spincharge})
\begin{eqnarray}
  H_c+H_{cs}&=&\int \frac{dx}{2\pi} \left[u_c (\pi \Pi)^2 +  u_c(\partial_x \phi)^2\right] + \frac{h}{\pi} \int dx \partial_x \phi   \nonumber \\
& &-\frac{2g_3}{(2\pi \alpha)^2} \int dx \cos 2 \phi   \nonumber  \\
& &+ \frac{2\Omega  \langle \cos \sqrt{2} \phi_s\rangle }{\pi \alpha} \int dx \cos 2 \theta,
  \label{eq:resc-ham}
\end{eqnarray}
and introduce the fermion operators
\begin{eqnarray}
  \label{eq:pseudofermions}
  \psi_R=\frac{e^{i(\theta-\phi)}}{\sqrt{2\pi \alpha}}, \\
  \psi_L=\frac{e^{i(\theta+\phi)}}{\sqrt{2\pi \alpha}},
\end{eqnarray}
to obtain~\cite{Giamarchi}
\begin{eqnarray}
  H_c+H_{cs}&=&-i u_c \int   dx (\psi^\dagger_R\partial_x  \psi_R -\psi^\dagger_L\partial_x  \psi_L) -i \frac U 2 \int dx (\psi^\dagger_R \psi_L - \psi^\dagger_L \psi_R ) \nonumber \\ 
& &  - h  \int   dx (\psi^\dagger_R \psi_R + \psi^\dagger_L\psi_L) \nonumber \\ 
& &  - i \frac{2\mu \Delta \langle \cos \sqrt{2}\phi_s\rangle} {\sqrt{t^2+\Delta^2}} \int dx (\psi^\dagger_R \psi^\dagger_L - \psi^\dagger_L \psi^\dagger_R)  .
  \label{eq:referm-ham}
\end{eqnarray}
The ground state of the Hamiltonian~(\ref{eq:referm-ham}) has been studied in \cite{orignac2017}. It has a phase transition point in the Ising universality class, whose order parameter is the BOW order parameter.
When $h\ne 0$, a disorder point~\cite{stephenson1970a,stephenson1970b} where charge density wave and BOW correlation functions remain short ranged but display incommensuration~\cite{orignac2017}  exists inside the superconducting phase.
The Fourier transforms of these correlation functions present a Lifshitz point~\cite{hornreich1975} where a double peak structure develops. In physical terms, the origin of the disorder point is simply that the doping induced by the chemical potential moves the Fermi wavevector away from $\frac \pi 2$ giving rise to incommensurate modulation of the density wave and bond order wave order parameters. While the fermion density displays no singularity at the disorder point, since $h$ can drive the Ising transition \cite{orignac2017}, it coupled to the Ising energy density operator. This implies that the fermion density has the same singularity at the Ising critical point as the energy density of the Ising model, \emph{i.e.} $\langle n_+ + n_- \rangle \sim (h-h_c) \ln |h-h_c|$.

\section*{References}


\end{document}